\documentclass[%
reprint,
 amsmath,amssymb,
 aps, physrev,
]{revtex4-2}

\usepackage{graphicx}
\usepackage{dcolumn}
\usepackage{bm}

\usepackage[utf8]{inputenc}
\usepackage[T1]{fontenc}

\usepackage{helvet}

\usepackage{caption}
\usepackage{capt-of}  

\usepackage[colorlinks=true,
            linkcolor=blue,    
            citecolor=blue,      
            urlcolor=blue         
           ]{hyperref}
\usepackage[dvipsnames]{xcolor}  

\begin{document}

\preprint{APS/123-QED}

\title{\textbf{Predicting  tunable nonreciprocal spin wave generation mediated by interfacial Dzyaloshinskii-Moriya interaction in magnonic heterostructures} 
}%

\author{Cameron A McEleney}
\affiliation{%
 James Watt School of Engineering, Electronics \& Nanoscale Engineering Division, University of Glasgow, Glasgow G12 8QQ, United Kingdom.
}%

\author{Karen L Livesey}
\affiliation{
School of Information and Physical Sciences, University of Newcastle, Callaghan NSW 2308, Australia
}%
\affiliation{Center for Magnetism and Magnetic Materials, Department of Physics and Energy Science,
University of Colorado at Colorado Springs, Colorado Springs, Colorado 80918, USA}

\author{Robert E Camley}
\affiliation{
Center for Magnetism and Magnetic Materials, Department of Physics and Energy Science,
University of Colorado at Colorado Springs, Colorado Springs, Colorado 80918, USA
}%

\author{Rair Mac\^edo}
\email{Contact author: Rair.Macedo@glasgow.ac.uk}
\affiliation{%
 James Watt School of Engineering, Electronics \& Nanoscale Engineering Division, University of Glasgow, Glasgow G12 8QQ, United Kingdom
}%



\begin{abstract}

Thin, metallic magnetic films can support nonreciprocal spin waves due to the interfacial Dzyaloshinskii-Moriya interaction (iDMI). However, these films typically have high damping, making spin wave propagation distances short (less than one micrometer). In this work, we theoretically study a thin ferromagnetic strip with iDMI and excite spin waves by driving a central segment of the strip. Spin waves propagate with different amplitudes to the left versus to the right from the driving region (i.e. nonreciprocity occurs) due to the iDMI. Our calculation based on spin-wave-dispersion plus our micromagnetic simulations both show that changing the driving segment width, driving frequency and static applied field strength tunes the nonreciprocity. Our calculation based on spin-wave-dispersion, using a so-called ``overlap function" will allow researchers to predict conditions of maximum nonreciprocity, without the need for computational solvers. Moreover, to circumvent the issue of short propagation distances, we propose a geometry where iDMI is only present in the driving region and low-damping materials comprise the remainder of the strip. Our calculations show significant spin wave amplitudes over several microns from the excitation region.

\end{abstract}

\maketitle


\section{\label{sec:Intro}Introduction}

Early studies on the impact of the antisymmetric Dzyaloshinskii-Moriya Interaction (DMI)~\cite{dzyaloshinsky1958,Moriya} concentrated on bulk materials where the influence of the DMI was weak, with a typical strength of less than 1\% of the symmetric Heisenberg exchange interaction. 
In contrast, the discovery of interfacial DMI (iDMI)~\cite{bogdanov2001chiral,bode2007},  which is much stronger, has ignited a significant set of research works in the last decade~\cite{hellman2017interface,kuepferling2023measuring,camley_2023_review}. 
This has led to studies of multiple skyrmion~\cite{roessler2006spontaneous} and skyrmion-like structures at room temperature~\cite{gobel2021beyond}, plus their motion caused by external fields or currents~\cite{fert2017magnetic,everschor2018perspective}.   

In addition, interfacial DMI can lead to significant changes to propagating spin waves. 
One promising effect involves the creation of nonreciprocal spin waves where the propagation wavevectors $+k$ and $-k$ have different frequencies, i.e. $\omega(k)\neq\omega(-k)$~\cite{udvardi2009chiral,zakeri2010asymmetric,Cortes-Ortuno_2013,Moon_2013,ma2014interfacial,kostylev2014interface,stashkevich2015experimental,nembach2015linear}. 
This nonreciprocal behavior has multiple practical applications and is used in isolators and circulators, for example \cite{kim2016spin,Bracher_2017,szulc2020spin}.
An important advantage of nonreciprocity caused by iDMI is that it occurs in very thin films, typically a few nanometers thick. This is in contrast to the nonreciprocal Damon-Eshbach mode which often requires films with thicknesses on the order of 1 micron~\cite{damon1961magnetostatic,camley1987nonreciprocal}.  


In this work, we explore the possibility of creating ultra-small devices-typical lateral sizes can be on the order of a few microns-based on magnonic heterostructures that support nonreciprocal spin-wave generation. 
A strip that is roughly 100~nm wide and 8~$\mu$m long is considered. 
The spin waves propagate along the strip, with the magnetization perpendicular to this direction but still in plane, and they are generated by an oscillating current carrying wire crossing the width of the strip. 
We change that driving field profile and use it to tune the nonreciprocity in a non-periodic heterostructure.
We find that the extent of the nonreciprocity between left- and right-traveling generated spin waves can be controlled by the width of the driving region (or antenna) and by the static applied magnetic field strength. That is, waves moving left from the driving region can be made to have a much larger amplitude than those moving right, or vice-versa.

The creation of unidirectional spin-wave emitters in systems with interfacial DMI has previously been investigated using analytical methods developed to describe and predict nonreciprocity~\cite{Bracher_2017}. 
Here, we provide a simpler formalism for understanding and predicting the nonreciprocity. This involves relating the width of the driving region to the wavelengths of the two spin waves ($-|k_L|$ and $+|k_R|$, respectively, for left and right) that are resonant with the driving frequency.  
We then demonstrate how this simplified model provides qualitatively similar results to those obtained with more complex models \cite{Bracher_2017} whilst also being in excellent agreement with our numerical experiments. This provides an easy predictive tool for identifying the combination of driving frequency, driving region width, and static applied field strength at which nonreciprocical spin wave generation is maximized. 

Finally, we note that earlier work showed that the distance propagated by the nonreciprocal waves away from the driving region was on the order of half a micrometer~\cite{jiang25}, a value somewhat too short for practical applications. This is because systems with appreciable iDMI typically also have large magnetic damping, due to the presence of a spin-orbit material interface~\cite{ourdani2021dependence}. We introduce a geometry where the iDMI is restricted to the driving region. This allows one to have a lower damping outside of the driving region (consistent with typical low-damping ferromagnetic metallic films), leading to longer propagation distances.

We begin by developing a simplified picture explaining the origin of the nonreciprocal magnetization dynamics, and perform a calculation based on linear spin wave dispersion to predict the conditions at which the strongest nonreciprocal spin wave generation emerges. Our model highlights the role of key parameters, including static applied field strength, driving field frequency, and the width of the signal line generating the spin waves. We then present micromagnetic results for a quasi-one-dimensional (1D) thin film strip structure based on an integration of the Landau-Lifshitz equations with demagnetizing factors appropriate to a strip structure. The calculation based on spin wave dispersion agrees qualitatively with the 1D micromagnetic simulations. Moreover, our 1D micromagnetics model is shown to give similar results to those produced by standard micromagnetic solver such as OOMMF or MuMax.

Micromagnetic calculations using OOMMF were done by Ma and Zhou~\cite{ma2014interfacial} in a bi-component magnonic waveguide, using a driving field in the center of the strip with a sinc function profile.
However, this did not address the possibility or advantages of a localized iDMI region. Recent work has looked at creating magnonic crystals by alternating wires with and without iDMI, to create flat magnonic bands.~\cite{flores2024selective} In that work, analytic examination of dispersion relations also proved useful in predicting complicated spin wave behavior in structured materials.

In Sec.~\ref{sec:Theory}, linear spin wave theory is presented so that our subsequent calculations can be understood. In Sec.~\ref{sec:Results}, the simulations are presented and the results for driven magnetization dynamics are shown. We explore tunabilty of the nonreciprocity and the prediction of maximum reciprocity without the need for simulations. Sec.~\ref{sec:concl} contains the conclusions and future outlook.

\section{\label{sec:Theory}Background: linear spin wave theory}

\begin{figure}
\includegraphics[width=0.48\textwidth]{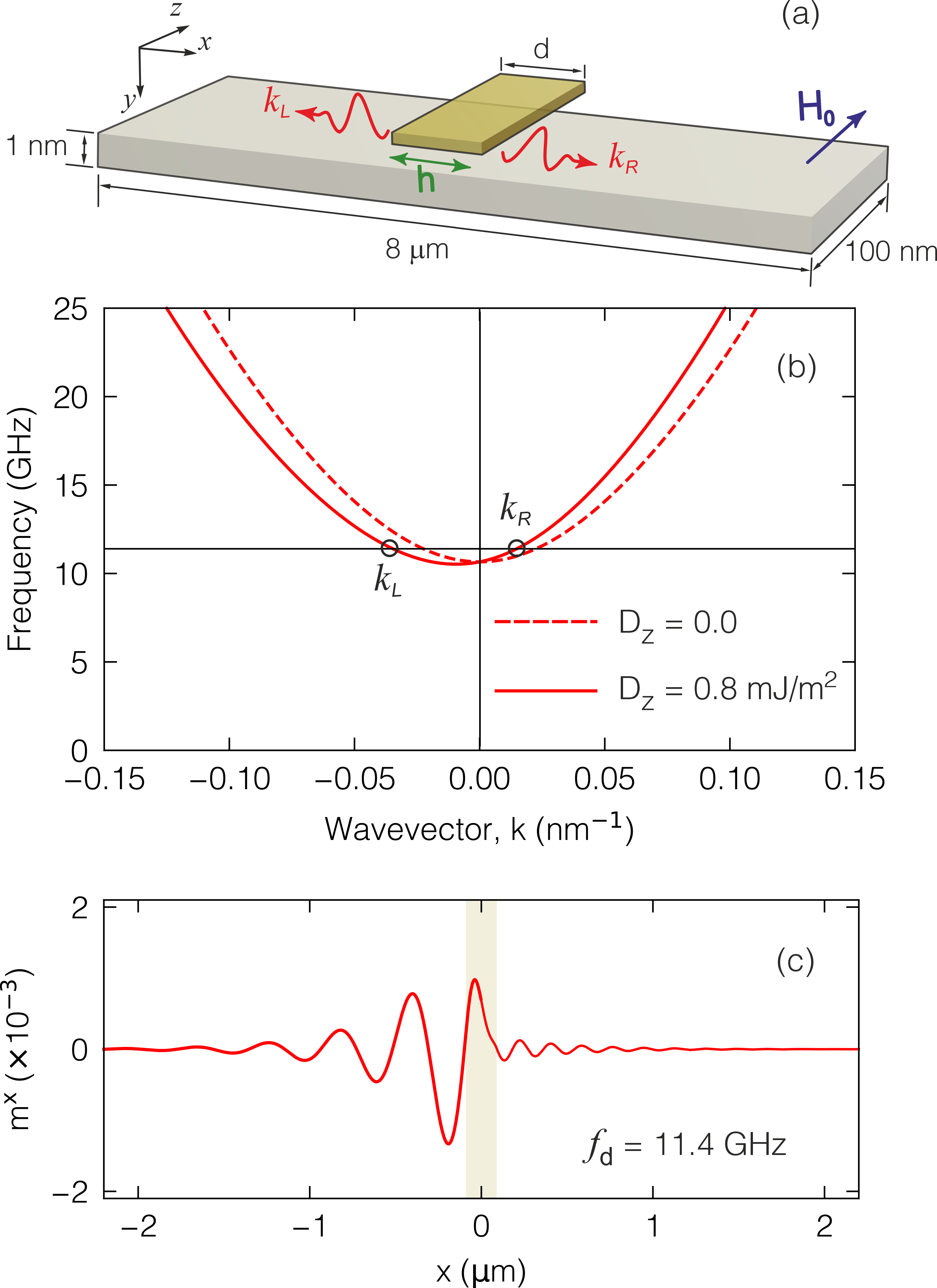}
\caption{\label{fig:Dispersion} (a) Schematic diagram of the thin film strip (modeled as a quasi-one-dimensional spin chain) with bias magnetic field applied along $z$ and the oscillatory driving field $\mathbf{h}$ along $x$. 
(b) Dispersion relation for the system in the presence of DMI (solid line) and when no DMI is present (dashed line).
(c) Snapshot of the magnetization ($m^x$) along the $x$ axis after 8~ns of driving for $D_z= $0.8~mJm$^{-2}$ and damping $\alpha=0.01$ across the full chain. The center of the spin-chain is driven ($d=$~180~nm, shaded region) at a driving frequency $f_d=$11.4~GHz and driving amplitude $h=$~0.03~mT.
The wave vector for left- and right-bound propagating spin waves at $f_d$ are indicated in (b) as $k_L$ and $k_R$, respectively. 
}
\end{figure}

Here, we study a ferromagnetic system with DMI, using a 1D spin chain model that represents propagation along the long axis ($x$ direction) in a thin film strip. (The strip is 80 times longer in $x$ than it is wide in $z$.)
In our model, we consider a static magnetic field $\mathbf{H_0}$ that is applied along the shorter in-plane $z$ direction as is illustrated in  Fig.~\ref{fig:Dispersion}(a). This is the Damon-Eshbach geometry, which gives the maximum effect of iDMI on the linear spin wave dispersion~\cite{camley_2023_review}.

For the chain of $N$ spins indexed by integer $i$ and with normalized magnetic moments $\mathbf{m_i}$, their dynamics is studied by time integration of the Landau–Lifshitz equation: 
\begin{equation}
\frac{\partial\mathbf{m_i}}{\partial t} = -|\gamma|\mu_0(\mathbf{m_i}\times\mathbf{H_i}),
    \label{LL}
\end{equation}
where $\gamma$ is the gyromagnetic ratio, $\mu_0$ is the permeability of free space, and $\mathbf{H_i}$ is the effective field at a site $i$. 
In a micromagnetics model, the effective field can be written as:   
\begin{equation}
\begin{aligned}
\mathbf{H_i}&=H_0\mathbf{\hat{z}}-M_s (N_x m_i^x\mathbf{\hat{x}} +N_y m_i^y\mathbf{\hat{y}} + N_z m_i^z\mathbf{\hat{z}})\\[8pt]
&+\frac{2A}{\mu_0 M_s}\frac{\partial^2\mathbf{m}(x)}{\partial x^2} + \frac{D_z}{\mu_0 M_s}\left(-\frac{\partial m^y}{\partial x}\mathbf{\hat{x}}+\frac{\partial m^x}{\partial x}\mathbf{\hat{y}}\right) ,
\end{aligned}
\label{field}
\end{equation}
where the first term is the static magnetic field with magnitude $H_0$ which here we take to be applied along $z$, the second term provides the effective demagnetizing field for a quasi-1D strip extended in the $x$ direction, the third term is the exchange field, and the last term is the DMI field. We note that, in the 1D chain, $y$ is the out-of-plane symmetry breaking direction. For spin waves propagating along $x$, a DMI vector $\mathbf{D} = D_z \mathbf{\hat{z}}$ is relevant. In Appendices ~\ref{App_A} and ~\ref{App_B}, respectively, the exchange and DMI fields in Eq.~\eqref{field} are converted into a form that is appropriate for discrete sites $i$ using finite difference expressions for the spatial derivatives.

For propagation perpendicular to $H_0 \mathbf{\hat{z}}$, the linear dispersion relation for spin waves can be found by making a few assumptions. Firstly, we assume the transverse components of $\mathbf{m_i}$ to vary as $m_i^{(x/y)}(x,t)\approx m_{i0}^{(x/y)} e^{i(k x-\omega t)}$  where $\omega$ is the angular frequency and $k$ is the wave number. 
We also make the standard linearization assumptions that the precession angle is small so that $m_i^z\approx1$, meaning that $m_{i0}^{(x/y)}$ are small quantities. 
From this, we arrive at
\begin{widetext}
\begin{align}
\frac{\omega}{|\gamma|} =\sqrt{\left[\mu_0 H_0+\frac{2A}{M_s}  k^2+M_s (N_x-N_z )\right]\left[\mu_0 H_0+\frac{2A}{M_s}k^2+M_s (N_y-N_z )\right] }+\frac{D_z}{M_s}k,
\label{disp}
\end{align}
\end{widetext}
which is the well-known result of Moon \emph{et al.}~\cite{Moon_2013} in the long-wavelength and magnetostatic limit. This is appropriate for very thin films (1~nm thick, as considered here) and for the small wavevectors accessible in Brillouin Light Scattering experiments.

In Fig.~\ref{fig:Dispersion}(b) we show the dispersion relation in Eq.~\eqref{disp} using parameters taken from Moon \emph{et al.}~\cite{Moon_2013}; namely, $\gamma/2\pi=28$~GHz/T, $M_s=$~800~kA/m, and the exchange stiffness constant $A=$~1.3$\times$10$^{-11}$~J/m. 
We take $D_z=$~0.8~mJ/m$^{2}$, which is lower than the value given in \cite{Moon_2013} but consistent with various experimental reports on thin film structures \cite{camley_2023_review}. 
We also take demagnetizing constants as $N_x\approx$~0.0, $N_y\approx$~0.9, and $N_z\approx$~0.1 which correspond to a thin film of dimensions 8~$\mu$m$\times$1~nm$\times$100~nm (calculated from \cite{Demag_Calculator}) and an applied field of $B_0=\mu_0 H_0=$~150~mT. We can see the classical asymmetrical behavior of the dispersion relation with respect to the direction of the wave vector, $k$ when DMI is nonzero (solid line).

We will return to these dispersion relation results in the next section, when developing an argument for how nonreciprocity can be maximized.

\section{\label{sec:Results}Driven spin wave results}

Now we go beyond the analytic calculation of spin wave frequencies and examine the magnetization dynamics that propagate outwards from a small region of the strip that is driven by an external rf field. One predicts that the waves propagating left (-$x$) and right ($+x$) will have a different amplitude due to DMI and that this may depend on the particular driving frequency $f_d$.

\subsection{Numerical micromagnetics calculation}

We numerically solve Eq.~\eqref{LL} forward in time with the addition of a damping term
\begin{equation}
    -|\gamma|\mu_0\alpha\left[\mathbf{m_i}\times(\mathbf{m_i}\times\mathbf{H_i})\right],
\end{equation}
where $\alpha=0.01$ is the Gilbert damping parameter at all sites $i$.
We also add an oscillatory driving field $\mathbf{h}(x,t)$ in the $x$ direction, spatially localized to the center of the chain, to Eq.~\eqref{field} for the effective field. This field turns on at time $t=0$ and is given by 
\begin{equation}
    \mathbf{h}(x,t)=g(x) ~h~ \cos (2\pi f_d t)\mathbf{\hat{x}},
\end{equation} 
where $f_d$ is the rf driving field frequency (varied) and $h=$~0.03~mT is the amplitude (fixed throughout this work). Here, $g(x)$ is taken to be a square driving profile on the order of $d=100-200$~nm wide, namely
\begin{equation}
    g(x) = \left\{
\begin{array}{ll}
1, &  -d/2 < x < d/2 \\
0, & \textrm{elsewhere}
\end{array} .
    \right.
    \nonumber
\end{equation}

Numerical integration was performed using a second-order Runge-Kutta scheme with time steps of 4$\times$10$^{-6}$~ns. This is a similar scheme as we used in an earlier study that did not include iDMI~\cite{macedo_2022}.
The system was driven uniformly along a block of 72 sites (or $d = $~180~nm, as we assume a micromagnetic cell size of 2.5~nm) located at the center of the chain and the resulting dynamics were recorded. 
Typical results illustrating the nonreciprocal behavior are shown in Fig.~\ref{fig:Dispersion}(c). 
The snapshot of the magnetization component $m^x$ as a function of position $x$ is presented after the system is driven for $t= $~8~ns (91 field cycles). 
This shows how, for a driving frequency $f_d=$~11.4~GHz, the amplitude is lower for right-bound spin waves, compared to left bound waves.  Nonetheless, it this case waves propagating in both directions die out after traveling about a micron.

\subsection{Overlap function to predict nonreciprocal propagation}

The difference in amplitude is a direct consequence of matching the width of the driving region $d$ to the two resonant spin waves' wavelengths. Namely, if the drive width $d$ is an integer multiple of the wavelength for a spin wave that is resonant with the driving field frequency, this will yield no propagation as there is no net magnetic moment to be excited.  Similarly, if the width of the driving region is a half-integer multiple of a resonant spin wave's wavelength, then maximal coupling will take place)~\cite{macedo_2022}. The key point is that, with DMI present, there are \emph{two} different wavelengths that are resonant with a given driving field frequency $f_d$, as seen in Fig.~\ref{fig:Dispersion}(b) and labeled by $k_L$ and $k_R$. 
Ideally, to get the largest nonreciprocity one wants the driving region width to be an integer multiple of one wavelength and a half-integer multiple of the other wavelength. 
In practice, significant nonreciprocity can be obtained simply by requiring that the driving region width $d$ is an integer multiple of the one of the resonant wavelengths. 

\begin{figure}
\includegraphics[width=0.45\textwidth]{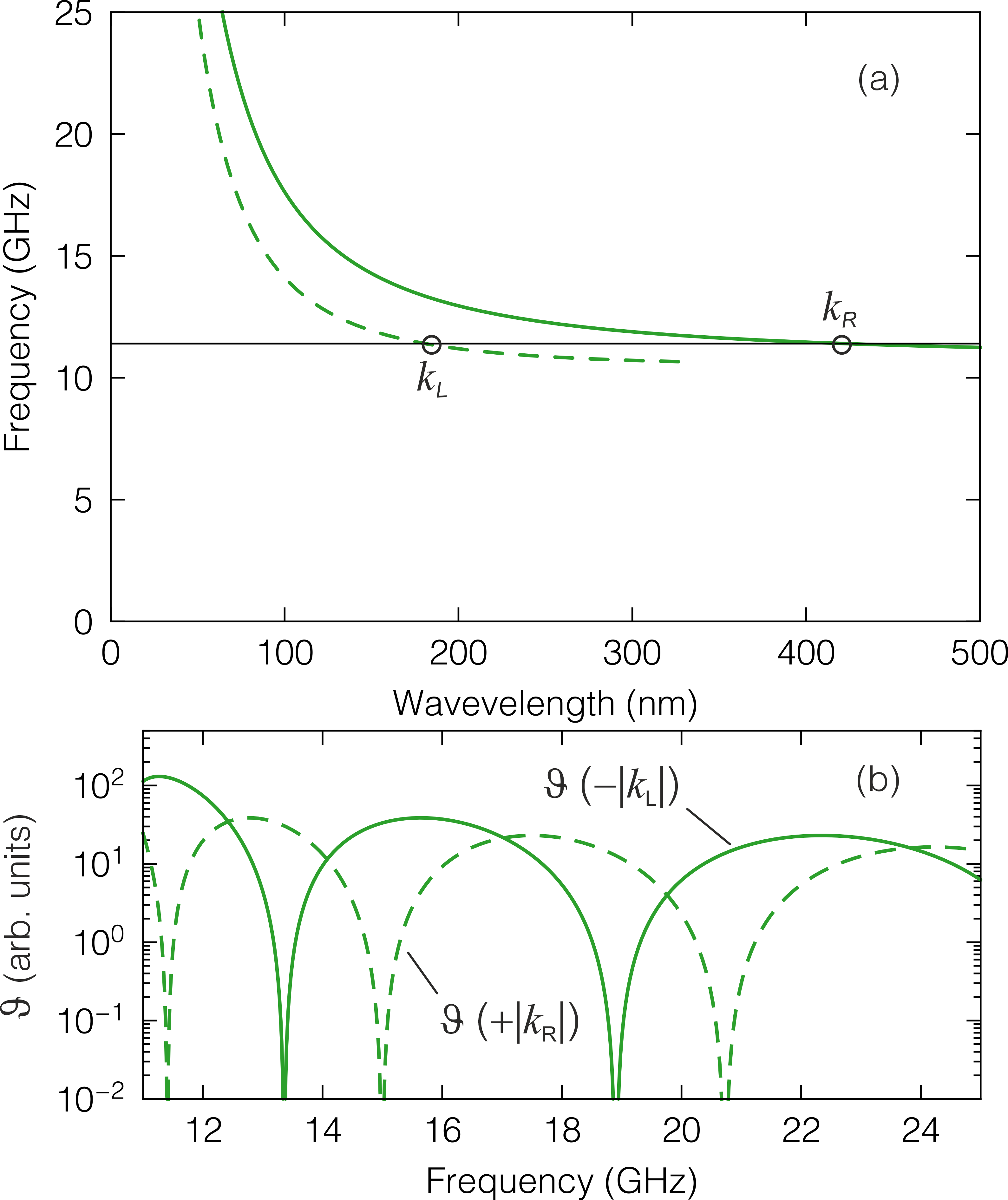}
\caption{\label{fig:wavelength} (a) Resonant spin wave frequencies as a function of wavelength for left (dashed) and right (solid line) bound propagation, obtained from the dispersion relation in Fig.~\ref{fig:Dispersion}(b). The horizontal line shows the case of $f_d=11.4$~GHz that is used throughout this work, with wavelengths corresponding to $k_L$ and $k_R$ labeled. (b) Equivalent overlap function $\vartheta$ as function of driving frequency for a driven region of width $d =$~180~nm, plotted on a log scale. 
}
\end{figure}

With this insight in mind, in the following we derive an ``overlap function" to predict the relative driving efficiency and the nonreciprocity possible. Rather than treating the dispersion relation in the usual way-finding frequency as a function of $k$ as done in Fig. 1(b)-we instead numerically invert Eq.~\eqref{disp} to find the two $k$ values resonant with a given frequency. From these $k$-values, we can obtain the corresponding wavelengths, as shown in Fig.~\ref{fig:wavelength}(a), which again shows a dramatic asymmetry between propagation in positive (dashed) and  negative (solid line) $x$ direction. 
This is particularly evident when comparing $k_L$ and $k_R$ for the same frequency $f_d$ (horizontal line), which, as introduced in Fig.~\ref{fig:Dispersion}, are highly nonreciprocal and therefore yield significantly different wavelengths.

We then assume that each linear spin wave mode within the DMI region can be modeled as a simple $\sin(kx)$ where $x$ is the spatial position and $k = k_L$ or $k=k_R$ is a given wavevector. By summing over the driving region (i.e., over $g(x)$), we define the overlap function $\vartheta$ of a spin wave $k$ resonant with the driving frequency and the driving field profile of width $d$:
\begin{equation}
    \vartheta(k)\equiv \int^{x=\infty}_{x=-\infty} g(x)~\sin(kx) dx =\int^{x=d/2}_{x=-d/2} \sin(kx) dx.
\end{equation}
Note that the units of the overlap function $\vartheta$ are those of length but have no real meaning. The relative size of the overlap is what will be of interest.

We note that there is a hidden assumption in this simplified picture, i.e. that you can specify the phase of the wave throughout the driving region for both $+|k_R|$ and $-|k_L|$. The phase of the spin wave with DMI may in fact not be uniform in this region. Previous work has shown that systems with DMI do not support regular standing waves with constant phase but instead have stationary nodes and traveling antinodes.~\cite{zingsem2019unusual,flores2020semianalytical} Nonetheless, as we will see, the assumption of uniform phase does not significantly change the general behavior and we are still able to infer how the diving field and spin wave mode overlap to predict strong nonreciprocity. 

The relative magnitude of the overlap function $\vartheta(k)$ then provides a prediction for which combinations of the driving frequency $f_d$ and driving region width $d$ will yield strong nonreciprocity. 

The results of this overlap function analysis are shown in Fig.~\ref{fig:wavelength}(b) for both the positive and negative $k$ branches extracted from the dispersion relation shown in Fig.~\ref{fig:Dispersion}(b).
We can see that there are various frequencies where a minimum driving efficiency---or overlap function---for one $k$ corresponds to a large value in the overlap function for $k$ in the opposite direction. For example, the first strong nonreciprocal frequency is 11.4~GHz, showing a minimum overlap between the spin wave mode and a driving region that is $d=180$~nm wide for $+|k_R|$, while a strong overlap is observed for $-|k_L|$. This argument involving an overlap function confirms the behavior presented in Fig.~\ref{fig:Dispersion}(c), where the left propagating wave has a much larger amplitude than the right propagating wave.

\subsection{Increasing the propagation distance}

\begin{figure}
\includegraphics[width=0.48\textwidth]{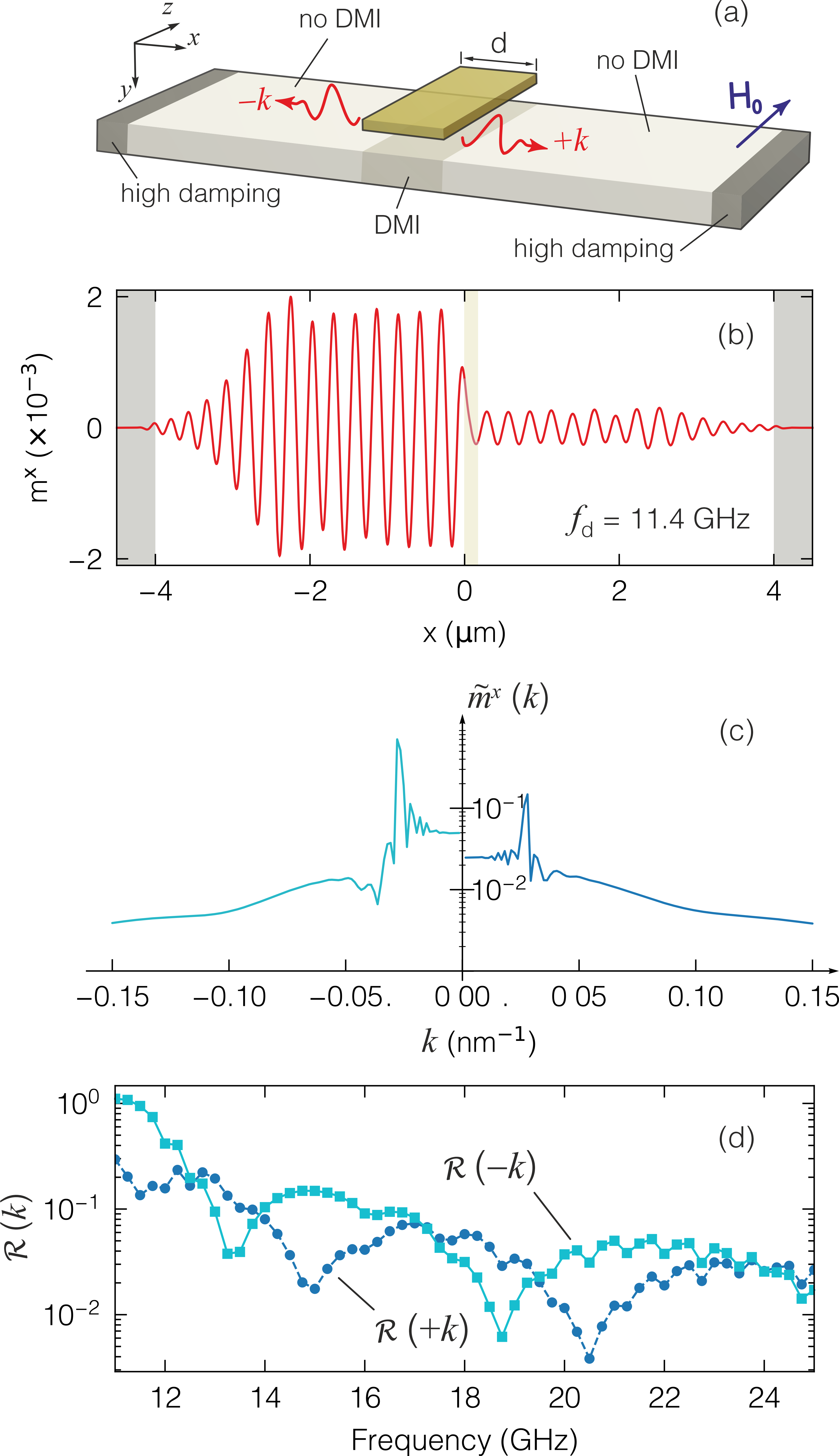}
\caption{\label{fig:theta} (a) Schematic diagram of the thin film strip (modeled as a quasi-one-dimensional spin chain) with bias magnetic field applied along $z$ and a drive region of width $d$, which is the only region with DMI. We assume that within the driving region, $D_z=$~0.8~mJm$^{-2}$ and $\alpha=0.01$. Elsewhere, $\alpha=$~10$^{-4}$. In the gray shaded regions at the ends, $\alpha$ linearly increases from 10$^{-4}$ to 0.2 at the very edges. 
(b) Snapshot of magnetization motion along the $x$ axis for $f_d=$~11.4~GHz after driving for 8~ns (91 cycles). 
(c) Spatial FFT of the data in Fig.~\ref{fig:theta}(b), termed $\tilde{m}^x(k)$. The dark blue line is obtained by taking the data to the right of the driving region ($+k$), while the cyan line is obtained by taking the data to the left of the drive ($-k$). 
(d) The maximum FFT value of $\tilde{m}^x(k)$, termed $\mathcal{R}(\pm k)$, for generated spin waves (somewhat equivalent to the overlap function). The dark, blue circles are for right- $(x>0)$ and the cyan squares are for left-propagation $(x<0)$. For clarity, the amplitudes are given on a log scale.
}
\end{figure}

Having now understood and contextualized creating nonreciprocal spin waves in a systems with DMI, we turn to one of the central aims of this work: optimizing nonreciprocity in a practical device. As seen in Fig.~\ref{fig:Dispersion}(c), one of the main problems of spin wave devices in DMI systems is the short-lived propagation due to the high damping of the structure where DMI has been typically observed. To circumvent that, we study an engineered heterostructure comprised of a system with DMI in the center, and no DMI elsewhere, as depicted in Fig.~\ref{fig:theta}(a). Such a structure could be easily constructed simply by depositing Pt, as an example, on top of the ferromagnetic thin film in the driving region only. In this way, the damping can be lower outside the driving region. Inside the driven, DMI region, we assume $\alpha=0.01$, which is consistent with DMI systems~\cite{ourdani2021dependence}. Elsewhere, $\alpha=$~10$^{-4}$ which is consistent with low damping ferromagnetic films.  

Note that in simulations we make the damping large at the two ends of the strip to minimize reflections which could complicate interpretation of the nonreciprocity. The gray shaded regions at either end in Fig.~\ref{fig:theta}(a), about 500~nm wide, are highly damped with $\alpha$ linearly increasing from 10$^{-4}$ to 0.2 at the very edges. 

This system is advantageous, as by using a simple ferromagnetic film with low damping outside of the driving region one can obtain longer spin wave propagation lengths. This is verified in the numerical experiment with a snapshot of the magnetization at time $t=8$~ns shown in Fig.~\ref{fig:theta}(b), where we repeat the driving mechanism shown in Fig.~\ref{fig:Dispersion}(c). Again, one can see a similar asymmetry in propagation; namely high amplitude propagation to the left and small propagation to the right. However, waves travel much further than in the previous example, reaching well beyond 10~$\mu$m from the excitation/driving region.

\subsection{Tuning the nonreciprocity via rf field frequency}

Now that the propagation distance is increased, we wish to compare the predictions of the overlap function (Fig.~\ref{fig:wavelength}(b)) for frequencies of maximum nonreciprocity to the results seen in our 1D micromagnetics simulations. 
In the ferromagnetic region without DMI, we need a measure of the response. 
We do this by taking the data from Fig.~\ref{fig:theta}(b) (or similar data at other frequencies) and doing a spatial fast Fourier Transform (FFT) of $m^{x}(x)$ on the left and right side of the driving region.
For consistency, we use $t=$~5~ns for all results we will show, but the response is not strongly dependent on time provided that the signal is allowed to propagate enough along the system.

As an example, we show the spatial FFT, $\tilde{m}^x(k)$ from Fig.~\ref{fig:theta}(b) in Fig.~\ref{fig:theta}(c). There are two appropriate peaks corresponding to waves moving to the right or left of the driving region (i.e. $\pm k$).
The $k$ values where the peaks occur are consistent with the dispersion relation for a simple ferromagnet at the driving frequencies. 
We therefore define a response function $\mathcal{R}(\pm k)$ as the maximum value of the FFT at the position $\pm k$.

In Fig.~\ref{fig:theta}(d) we plot the equivalent response function $\mathcal{R}(\pm k)$ as a function of the driving frequency. For this, we take data similar to that given in Fig.~\ref{fig:theta}(b) and take spatial FFTs for either the left (teal squares) or right (dark blue circles) side of the signal, then repeat this for a range of frequencies. 
Fig.~\ref{fig:theta}(d) can be compared to the plot of the overlap function shown in Fig.~\ref{fig:wavelength}(b). Specifically, the frequencies at which maximum non-reciprocity takes place---that is, maxima and minima coincide---show excellent agreement with our simple overlap function derived from the linear spin wave dispersion relation. As an example, the maximum/minimum near 15~GHz lines up in both plots.

Note that the left- and right-moving waves considered in Fig.~\ref{fig:theta}(d) now have the same magnitude, unlike in Fig.~\ref{fig:wavelength}.
In contrast to the situation with DMI throughout the entire strip (Sec.~\ref{sec:Results}~B), the absence of DMI outside the driving region means that the \emph{same} wavenumber magnitude corresponds to the maximum in the spatial FFT, both left and right of the driving region. The DMI in the driving region introduces nonreciprocal spin wave amplitudes, but the wavelengths are now the same on the left and right, as can be seen looking at Fig.~\ref{fig:theta}(b).

To quantify the degree of nonreciprocity in the spin wave amplitude, we use the dimensionless parameter $\eta$. For the numerical micromagnetic simulations, $\eta_\textrm{num}$ is defined at a particular driving frequency as the natural logarithm of the ratio between the maximum amplitudes of the FFT of the dynamic magnetization component for right- and left-propagating spin waves or simply:
\begin{equation}
    \eta_{\textrm{num}} = \ln\left[\frac{\mathcal{R}(+k)}{\mathcal{R}(-k)}\right].
\end{equation}

For the simple model based on the spin wave dispersion, the related $\eta_{\textrm{simple}}$ is obtained from the related overlap function, which quantifies the coupling efficiency between the excitation and spin-wave modes, namely:
\begin{equation}
    \eta_{\textrm{simple}} = \ln\left[\frac{\vartheta(+|k_R|)}{\vartheta(-|k_L|)}\right].
\end{equation}
Positive (negative) values of $\eta$ indicate stronger propagation for right- (left-) bound waves, while the absolute magnitude $|\eta|$ (i.e. peaks and dips) represents the strength of the nonreciprocity, which we can clearly see varying strongly with frequency of the driving field.

\begin{figure}
\includegraphics[width=0.45\textwidth]{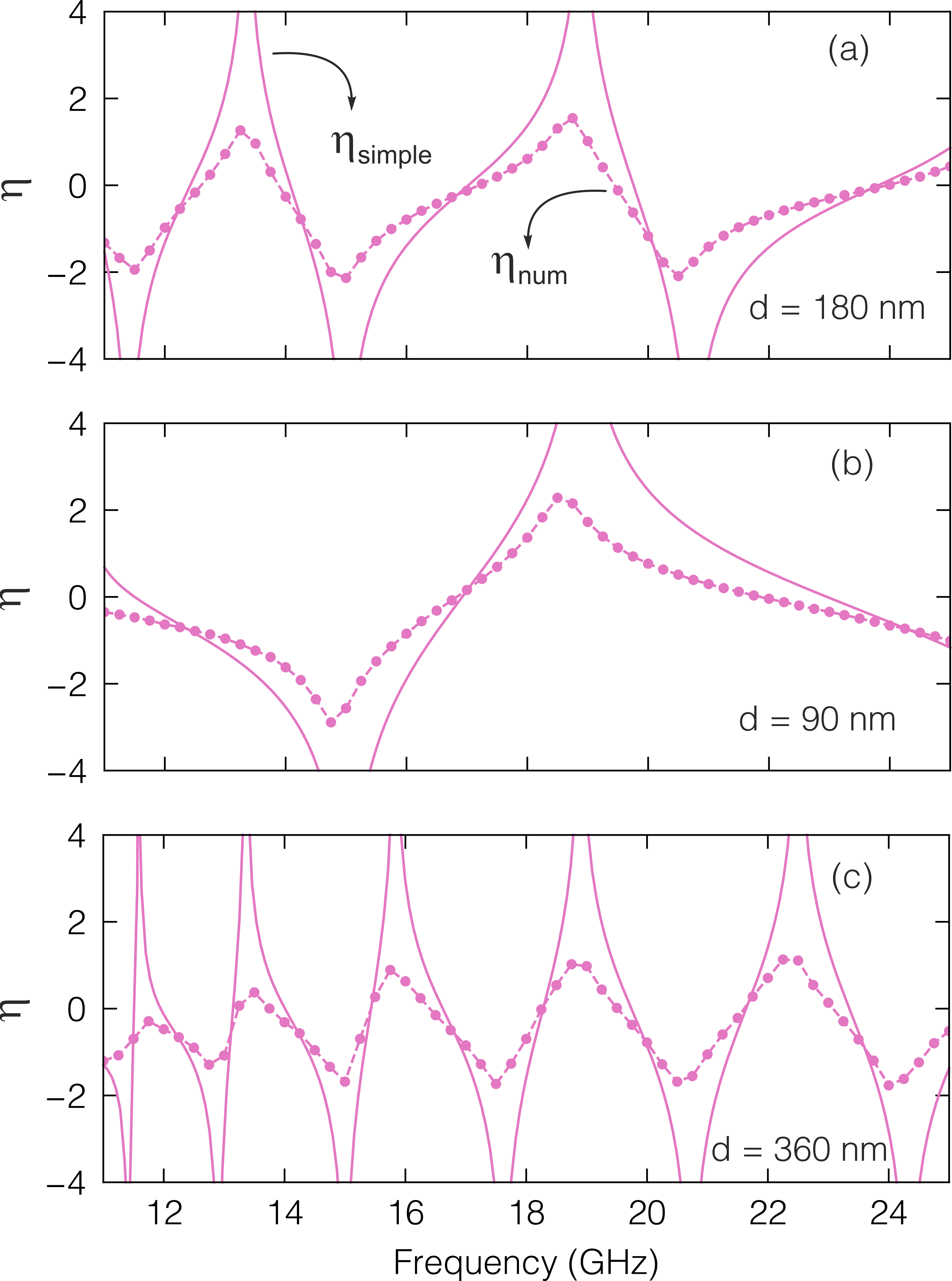}
\caption{\label{fig:eta} Nonreciprocity efficiency constant $\eta$ as a function of driving field frequency $f_d$. We compare the micromagnetic solutions $\eta_{\textrm{num}}$ (symbols) with the prediction based on spin wave dispersion $\eta_{\textrm{simple}}$ (solid lines). These are given for three different driving region widths (a) $d=$~180~nm, (b) $d=$~90~nm, and (c) $d=$~360~nm. 
}
\end{figure}

In Fig.~\ref{fig:eta}(a) the numerical (dots) and simple estimate (solid line) for the nonreciprocity measure $\eta$ is plotted as a function of driving frequency, with all parameters the same as considered in Fig.~\ref{fig:theta}. One sees that the two methods predict the same frequencies for extreme nonreciprocity, for example at $f_d=11.4$~GHz. This means that the simple method can be used by researchers to engineer magnetic heterostructures for nonreciprocal spin wave applications, without the need for long numerical simulations.

To highlight how this behavior can be further tuned, we vary the width of the diving region making it shorter $d= $~90~nm, and longer $d= $~360~nm, with results given in Figs.~\ref{fig:eta}(b) and (c), respectively. The dotted lines are again nonreciprocity efficiency from the micromagnetic simulations as a function of frequency for the system with nonuniform damping. The solid lines are results from the overlap function. We can clearly see how a shorter drive region (Fig.~\ref{fig:eta}(b)) yields only two regions of strong non-reciprocity up to 25~GHz: one for right- (near 18~GHz) and one for left-propagation (near 15~GHz). A longer driving region (Fig.~\ref{fig:eta}(c)) yields many nonreciprocal frequencies. Therefore, the width of the driving region can be used to either tune nonreciprocity to a specific frequency, or to isolate nonreciprocity to only a certain frequency across a wide-band. This behavior is similar to that shown in Ref.~\cite{Bracher_2017} where similarly distinct maxima and minima were predicted due to the splitting of the dispersion relation in combination with the wave-vector selective excitation by a finite drive.

\subsection{Tuning with static applied field strength}

\begin{figure}
\includegraphics[width=0.45\textwidth]{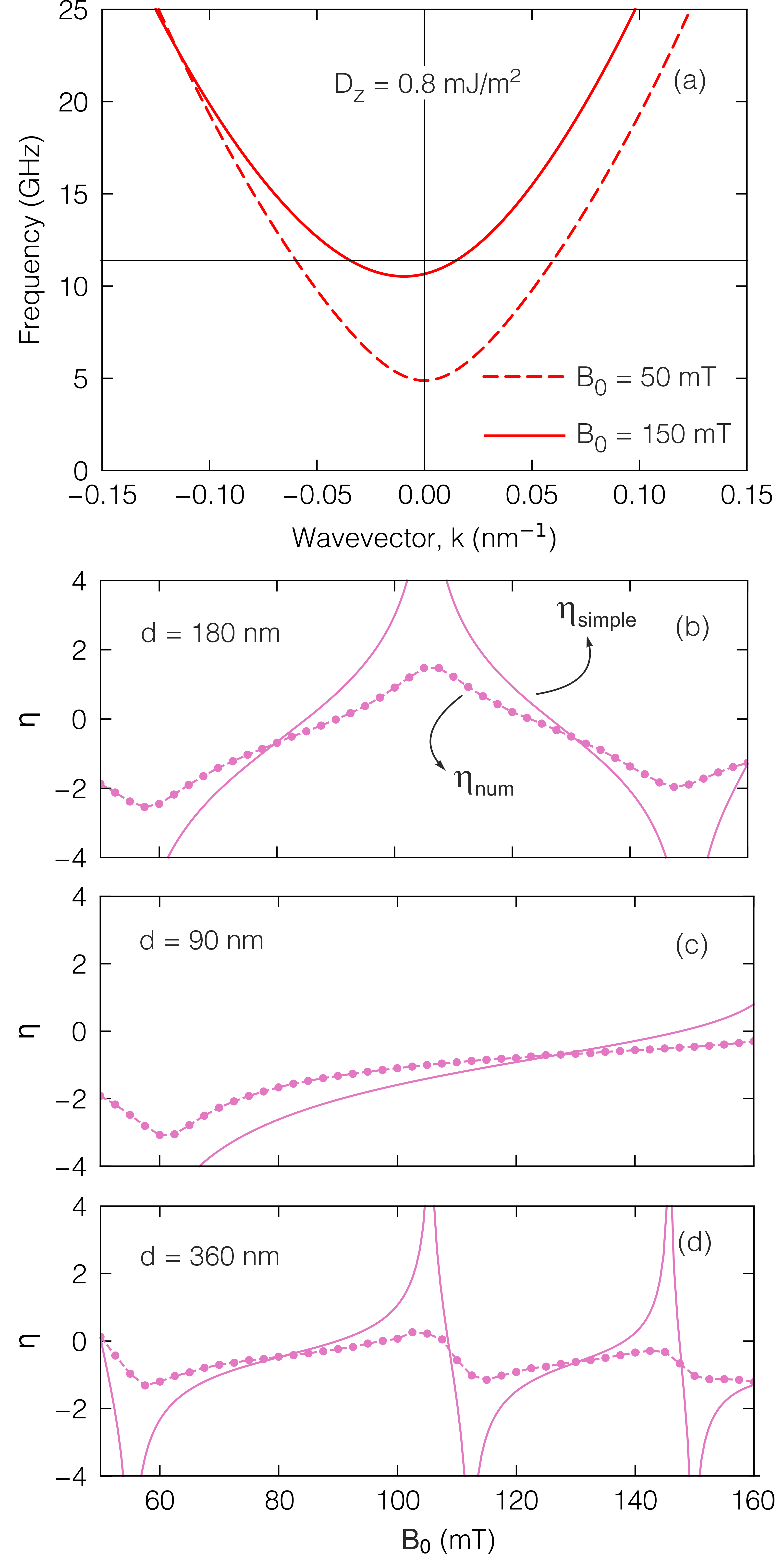}
\caption{\label{fig:FieldTune} (a) Dispersion relation for two applied fields of $B_0=$~50~mT (dashed line) and $B_0=$~150~mT (solid line). Both cases are for a thin strip, considering the presence of iDMI interaction. The nonreciprocity dimensionless parameter $\eta$ for both the numerical simulations ($\eta_{\textrm{num}}$) and predictions based on the dispersion relation ($\eta_{\textrm{simple}}$) are given for (b) $d=$~180~nm, (c) $d=$~90~nm and (d) $d=$~360~nm, all as function of applied field $B_0$. Through (b) to (d) the driving frequency is kept constant,  $f_d=$~11.4~GHz, and other parameters are the same as used in Fig.~\ref{fig:theta}. 
}
\end{figure}

While the width of the driving region (also the region with DMI) can be used to control the frequencies at which nonreciprocal driving of spin waves occurs, in practice this would not be efficient as it would require changing the magnonic-heterostructures for each experiment. However, the dispersion relation can be controlled by the strength of the static applied magnetic field. In Fig.~\ref{fig:FieldTune}(a) we show the dispersion relation for a system with DMI, for two applied field strengths $B_0= $50~mT (dashed line) and $B_0=$~150~mT (solid line). For the driving frequency used before ($f_d=$~11.4~GHz, horizontal line), the resonant wave vectors are substantially different for the two applied fields.

As a consequence, it is possible to use the applied field to control the nonreciprocity of excited waves in a similar fashion to what we have seen before when changing the driving frequency. In Fig.~\ref{fig:FieldTune}(b) we show the result for the FFT nonreciprocity parameter $\eta$ in a numerical experiment where we set the frequency ($f_d=$~11.4~GHz) and the driving region width ($d=180$~nm) and change the strength of the applied field. We can efficiently tune the nonreciprocity from suppressing right-bound spin waves (at around 60~mT) to suppressing left-bound waves as the field increases (peaking at about 110~mT). Notably, these fields are experimentally attainable.

As seen in the previous cases, this field tunability can be drastically modified if we change the width of the driving region. In Figs.~\ref{fig:FieldTune}(c) and (d), we show examples for the same applied field range and driving frequency of $f_d=$~11.4~GHz but now for $d= 90$~nm and 360 nm, respectively. For the shorter driving region, only one region of maximum nonreciprocal behavior is observed, while for the longer driving region there are three such regions within the field range displayed.

\section{\label{sec:concl}Discussion and Conclusions}

Our results have shown that engineered magnetic heterostructures can support nonreciprocal spin wave generation offering a variety of parameters with which one can control the nonreciprocal behavior. The spin waves can propagate over 3~microns in the engineered systems where iDMI and damping varies along the strip's length. In particular, the spin wave \emph{amplitude} is different to the left and right of the driving region in this situation and the dominant wavelength is the same.
We have also verified the results discussed here---for a 1D strip obtained using our own Fortran codes---by comparing them with MuMax and OOMMF simulations, and we find qualitatively good agreement.

Most excitingly, the conditions for nonreciprocal generation (driving frequency, static applied field, etc) can be accurately predicted by examining the analytic spin wave dispersion relation with iDMI, and constructing an overlap function. This gives scientists and engineers a tool for designing heterostructures without the need for micromagnetic simulations.

It is worth noting that these heterostructures, where the iDMI and the damping are different in different regions within the structure, can introduce potential issues that are not observed in continuous films. For instance, at discontinuities one can potentially observe reflections~\cite{verba2020spin} that may impact device performance. Here, we chose to focus on iDMI constants that are lower than typical values previously used \cite{Moon_2013} but still in line with experimentally reported values in various experiments \cite{camley_2023_review}, and in this case reflections at interfaces does not seem to be a significant issue.  However, as one goes to significantly higher iDMI values, this can become a problem for coherent spin wave propagation. This also has to be considered for wider driving regions where back and forward propagation inside the iDMI region can lead to interference that can be detrimental to the overall device operation (see Appendix~\ref{App_C} for additional calculations exemplifying this phenomenon).   
  
We have not considered the case where anisotropy may be present. We note this as anisotropy can be another interesting way of controlling nonreciprocity. For instance, consider the simplest case-—that of uniaxial anisotropy-—with a contribution added to the effective field in Eq.~\eqref{field} given by
\begin{equation}
    \mathbf{H^A_i} = +\frac{2K^u_i}{M_s}(\mathbf{m_i.\mathbf{e}^u})\mathbf{e}^u.
\end{equation}
Here, $K_i^u$ is the anisotropy constant and $\mathbf{e}^u$ is a unit vector pointing along the easy axis. If we set the anisotropy to be along $z$, then the anisotropy field written above provides a similar contribution to the static applied field $H_0$ in the dispersion relation. Therefore, it should be possible to obtain nonreciprocity without the need for an applied external field \cite{Kuanr03}.
We note that for a thin film geometry such as the one we investigated in this work, the crystalline anisotropy field would have to be larger than the in-plane shape anisotropy field.

We have demonstrated how nonreciprocity can be introduced, tuned, and—perhaps most importantly—sustained for longer lifetimes by engineering magnonic heterostructures that combine regions with iDMI and regions with symmetric, exchange-only interactions. This approach allows control over the spin-wave wavelength mismatch between counter-propagating modes, enabling selective enhancement or suppression of propagation. This provides a route towards highly controllable, long-lived nonreciprocal magnonic devices.

\begin{acknowledgments}
\noindent C.A.M. was supported by the Engineering and Physical Sciences Research Council (EPSRC) under grant number EP/S023321/1.
The work of R.E.C. was partially supported by the Australian-American Fulbright Commission through a Distinguished
Chair position at the University of Newcastle, Australia.
The School of Information and Physical Sciences at the
University of Newcastle---where a portion of this work was performed---is acknowledged for their travel support and hospitality to R.M. and R.E.C.

\noindent For the purpose of open access, the authors have applied a Creative Commons Attribution (CC BY) license to any Author Accepted Manuscript version arising from
this submission.
\end{acknowledgments}

\appendix

\section{\label{App_A}Micromagnetic effective fields (Exchange) }

As discussed in the main text, we consider a one-dimensional magnetic system extended along the $x$-axis. The magnetization is described in terms of the reduced magnetization vector 
\begin{equation}
    \mathbf{m}(x,t) = \frac{\mathbf{M}(x,t)}{M_s}
\end{equation}
where $M_s$ is the saturation magnetization and $|\mathbf{m}|=1$.

In a micromagnetic approximation, the exchange interaction is modeled using a phenomenological continuum description. The exchange energy density (per unit volume) for a 1D system can be written as
\begin{equation}
    \mathcal{E}^E=A\left|\frac{\partial \mathbf{m}}{\partial x}\right|^2,
\end{equation}
where the exchange stiffness $A$ is taken to be spatially uniform.

The corresponding exchange effective field is obtained by taking the functional derivative of the exchange energy with respect to the magnetization
\begin{equation}
    \mathbf{H^E}(x) = -\frac{1}{\mu_0M_s}\frac{\delta \mathcal{E}^E}{\delta \mathbf{m}(x)}=\frac{2A}{\mu_0M_s}\frac{\partial^2\mathbf{m}(x)}{\partial x^2}
\end{equation}
This field acts to minimize spatial variations of the magnetization, effectively favoring alignment of neighboring spins.

For numerical simulations of the LLG equation, we discretize the spatial derivatives using a finite-difference scheme on a regular lattice with spacing $\Delta x$. The second derivative is then approximated at site $i$ as
\begin{equation}
    \frac{\partial^2\mathbf{m}(x)}{\partial x^2}\approx\frac{\mathbf{m_{i+1}}-2\mathbf{m_{i}}+\mathbf{m_{i-1}}}{(\Delta x)^2},
\end{equation}
where $i+1$ and $i-1$ denote the two nearest neighbor micromagnetic spins in the $x$ direction.
The discrete form of the exchange field therefore reads
\begin{equation}
    \mathbf{H_i^E}=\frac{2A}{\mu_0M_s}\frac{1}{(\Delta x)^2}(\mathbf{m_{i+1}}-2\mathbf{m_{i}}+\mathbf{m_{i-1}}).
\end{equation}
We note that in the linear regime considered here, where the dynamics are dominated by small transverse deviations from the equilibrium magnetization, the term proportional to $-2\mathbf{m_i}$  does not contribute to the torque $\mathbf{m_i}\times\mathbf{H_i}$, and can therefore be omitted without affecting the spin-wave dynamics.

From Moon \emph{et al.} \cite{Moon_2013} we have $A=$~1.3$\times$10$^{-11}$~Jm$^{-1}$ and $M_s=$~8.0$\times$10$^5$~Am$^{-1}$. If we assume a micromagnetic cell size $\Delta x=$~2.5~nm, this gives:
\begin{equation*}
    \mu_0 H_E\approx 5.2~T.
\end{equation*}

\section{\label{App_B}Micromagnetic effective fields (iDMI)}

Similarly, we can model the iDMI interaction using a phenomenological continuum description. The iDMI energy density can be written as
\begin{equation}
\mathcal{E}^{DMI}=\mathbf{D}\cdot(\mathbf{m}\times\mathbf{\nabla m}).
\end{equation}
Since we are considering a 1D system extended along $x$, only the spatial variation or gradient along $x$ is considered ($\partial y=\partial z =0$). Therefore, the energy becomes
\begin{equation}
    \mathcal{E}^{DMI}=\mathbf{D}\cdot \left(\mathbf{m}\times\frac{\partial\mathbf{m}}{\partial x}\right).
\end{equation}

Here, we are also considering that the iDMI vector only has a component along the $z$ direction so that $\mathbf{D}=D_z \mathbf{\hat{z}}$. This gives
\begin{equation}
    \mathcal{E}^{DMI}=D_z \left(m^x\frac{\partial m^y}{\partial x}-m^y\frac{\partial m^x}{\partial x}\right).
\end{equation}

The effective field associated with the iDMI interaction, which enters the Landau Lifshitz equation, is defined as:
\begin{equation}
\begin{aligned}
    \mathbf{H^{DMI}}(x) &= -\frac{1}{\mu_0M_s}\frac{\delta \mathcal{E}^{DMI}}{\delta \mathbf{m}(x)}\\[8pt]
    &=\frac{D_z}{\mu_0M_s}\left(-\frac{\partial m^y}{\partial x}\mathbf{\hat{x}}+\frac{\partial m^x}{\partial x}\mathbf{\hat{y}}\right) .
    \end{aligned}
\end{equation}
This field drives chiral rotation of the magnetization in the $xy$-plane along $x$.

For numerical simulations of the Landau-Lifshitz equation, we discretize the spatial derivatives using a central difference scheme on a regular lattice with spacing $\Delta x$. The first derivatives are then approximated as
\begin{equation*}
\begin{aligned}
    \frac{\partial m^x}{\partial x}\approx \frac{m^x_{i+1}-m^x_{i-1}}{2 \Delta x} ,\\[8pt]
    \frac{\partial m^y}{\partial x}\approx \frac{m^y_{i+1}-m^y_{i-1}}{2 \Delta x}.
\end{aligned}
\end{equation*}

Substituting this into the continuous field we obtain the discrete form of the exchange field
\begin{equation}
\begin{aligned}
    \mathbf{H_i^{DMI}}= &\frac{D_z}{\mu_0M_s}\frac{1}{2\Delta x} \\[8pt]
    & \times\left[-\left(m^y_{i+1}-m^y_{i-1}\right)\mathbf{\hat{x}}+\left(m^x_{i+1}-m^x_{i-1}\right)\mathbf{\hat{y}}\right].
    \end{aligned}
\end{equation}
Here, $D_z$ is the effective micromagnetic iDMI constant (energy per unit length). 
We note that this can be related to the atomic iDMI energy by $D_z=2D_{ij}/a$, where $a$ is the distance between two neighboring atomic sites.~\cite{camley_2023_review}

If $D_z= $~0.8~mJm$^{-2}$ and $\Delta x=$~2.5~nm, then the effective iDMI field is
\begin{equation*}
    \mu_0 H_\text{DMI}\approx 0.2~T,
\end{equation*}    
which is roughly 4\% of the symmetric exchange field.

\section{\label{App_C}Boundary Effect }

\begin{figure}[h]
\includegraphics[width=0.45\textwidth]{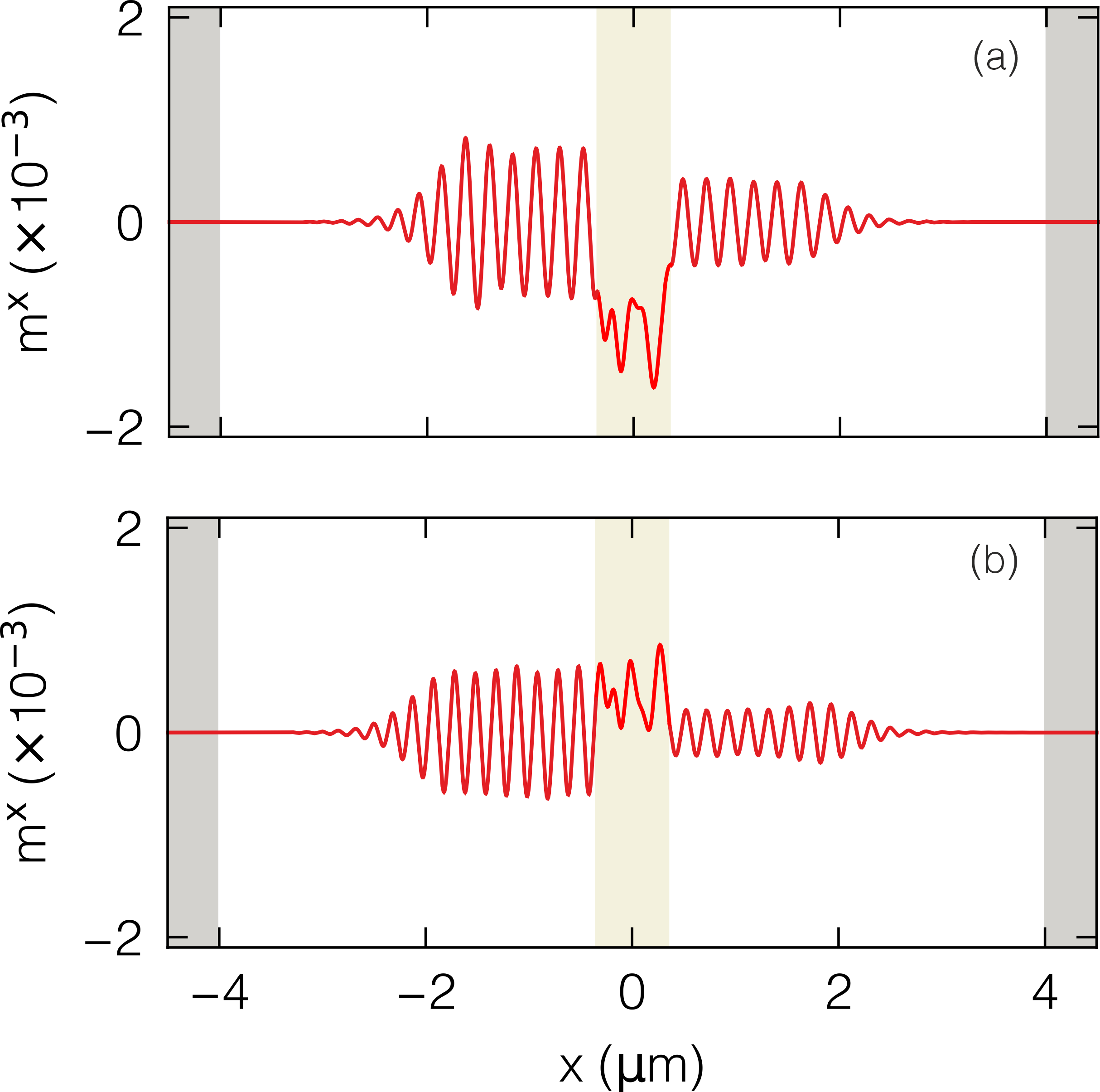}
\caption{\label{fig:App} Snapshot of spin-wave propagation along the $x$ axis for (a) $f_d=$~11.75~GHz and (b) $f_d=$~12.07~GHz to show the result of reflections within the driving region. 
We assume that within the driving region of width $d=$~0.72~$\mu$m (shaded in center), there is iDMI and higher damping.
}
\end{figure}

Throughout this work we have considered a magnonic heterostructure that contains abrupt interfaces between regions where iDMI is present and regions without iDMI. This abrupt change can introduce reflections at the interfaces. Snapshots are shown in Fig.~\ref{fig:App} for a larger driving region ($d=0.72~\mu$m) than considered in the main text, so the effect of reflections can be better observed. As in the main text, an iDMI interaction between the spins of $D_z=$~0.8~mJm$^{-2}$ is present in the central driving region, while no iDMI is present outside this region. 
We also take the damping to be different in both regions. 
In the iDMI region, the damping is $\alpha=$~10$^{-2}$, which is consistent with highly damped DiMI systems, and elsewhere it is taken to be $\alpha=$~10$^{-4}$ which is consistent with low damping ferromagnetic films.

In Fig.~\ref{fig:App}(a) we show the case where nonreciprocal generation is not expected to be dramatic ($f_d=11.75$~GHz). Within the driving region (shaded) reflections lead to a larger amplitude for magnetization motion than outside it. 

On the other hand, the effect is not as dramatic when the expected nonreciprocity is more pronounced (see Fig.~\ref{fig:App}(b) for $f_d = 12.07$~GHz, where right-bound spin wave propagation should be suppressed). However, reflections within the driving region lead to less selective nonreciprocity when compared to the case shown in Fig.~\ref{fig:theta}(b).


\bibliography{apssamp}

\end{document}